# Hard superconductivity of a soft metal in the quantum regime

MUSTAFA M. ÖZER[1], JAMES R. THOMPSON[1,2] AND HANNO H. WEITERING[1,2]

[1]*Department of Physics and Astronomy, The University of Tennessee, Knoxville, TN 37996, USA.* [2]*Condensed Matter Sciences Division, Oak Ridge National Laboratory, Oak Ridge, TN 37831, USA.*

**Superconductivity is inevitably suppressed in reduced dimensionality[1-9]. Questions of how thin superconducting wires or films can be before they lose their superconducting properties have important technological ramifications and go to the heart of understanding coherence and robustness of the superconducting state in quantum-confined geometries[1-9]. Here, we exploit quantum confinement of itinerant electrons in a soft metal to stabilize superconductors with lateral dimensions of the order of a few millimeters and vertical dimensions of only a few atomic layers[10]. These extremely thin superconductors show no indication of defect- or fluctuation-driven suppression of superconductivity and sustain supercurrents of up to 10% of the depairing current density. The extreme hardness of the critical state is attributed to quantum trapping of vortices. This study paints a conceptually appealing, elegant picture of a model nanoscale superconductor with calculable critical state properties.**

**It indicates the intriguing possibility of exploiting robust superconductivity at the nanoscale.**

Pb has the peculiar property that the wavelength of the highest occupied electron level matches the atom-layer spacing along the <111> crystal direction almost perfectly, so that two atomic layers of Pb accommodate one-and-a-half "Fermi wavelength". This electronic property can have profound consequences for the hetero-epitaxial growth of very thin Pb(111) films[10], provided that certain kinetic growth conditions are met[11-18]. Between 200 and 250 K, atomically flat crystalline Pb films on Si(111) substrates evolve in a quasi bilayer-by-bilayer fashion, beginning at 5 monolayers (ML)[11-18]. The bilayer growth is periodically interrupted by the growth of a single-atom layer or even a trilayer.[11] This remarkable growth mode can be attributed to strong Friedel oscillations in the electron density, in conjunction with the nearly perfect matching of lattice spacing and Fermi wavelength[11]. In this "quantum growth" regime, lattice strain is of minor importance because the metal is soft and quantum size effects are strong. The atomically flat Pb terraces are limited in size by the width of the underlying Si terraces[6,11]. This level of smoothness, which is astonishing for metal growth on semiconductors, provides the extraordinary opportunity to explore the superconductive properties of highly crystalline films with atomically controlled thickness.

Using contactless magnetic methods, we obtained the thermodynamic critical temperature $T_c$ and upper critical magnetic field $H_{c2}(T)$, shown in Fig. 1, and the non-equilibrium critical current density $J_c(T,H)$ for films with thickness $d$ = 5-18 ML. This layer count excludes the interfacial wetting layer, which is one atomic layer thick[15,18].

Figure 1 (inset) shows $T_c$ plotted as a function of $1/d$. The $T_c(d)$ data are perfectly linear in $1/d$ and extrapolate to the bulk $T_{c0} = 7.2$ K ($d = \infty$). Accordingly, $T_c(d) = T_{c0}(1 - d_c/d)$. The extrapolated threshold $d_c$ for the emergence of superconductivity is roughly 1.5 ML. Note that if the wetting layer is included in the layer count, $T_c(d)$ no longer extrapolates to the bulk $T_{C0}$. This suggests that the interfacial wetting layer does not strongly participate in the superconductivity. Our $T_c$ data do not exhibit quantum size effect oscillations, which are observable only in the classical layer-by-layer growth regime[6]. A $1/d$ dependence arises naturally from the inclusion of a surface energy term in the Ginzburg-Landau (GL) free-energy of a superconductor[19].

The fundamental length scales in the films are significantly affected by the two-dimensional (2D) geometry. For instance, the GL coherence length $\xi_{GL}(T)$ of the films is reduced significantly below the BCS value of bulk Pb, $\xi_0^{bulk} = 905$ Å. The in-plane $\xi_{GL}(T)$ was obtained from $H_{c2}(T)$ measurements via the relation $H_{c2}(T) = \Phi_0 / 2\pi \xi_{GL}^2(T)$ where $\Phi_0$ is the flux quantum[20]. Figure 1 shows $H_{c2}(T)$ of a few thin films. $H_{c2}(T)$ varies nearly linearly with temperature, conforming to the standard GL dependence[20] $\xi_{GL}(T) \propto (1 - T/T_c)^{-1/2}$. $H_{c2}(T)$ was extrapolated to 0 K using Bulaevskii's expression for 2D superconductors[21], giving $\xi_{GL}(0) \approx 230$ Å for a 9 ML Pb film. This significant reduction in coherence length, as compared to the BCS value $\xi_0^{bulk}$ of bulk Pb, is attributed primarily to the reduced electronic mean free path $l(d)$ in the films. Using the "full value" expression[20]

$$\xi_{GL}(T=0, d) \cong 0.739 \{\xi_0^{-2} + 0.882[\xi_0 \times l(d)]^{-1}\}^{-1/2}$$

one estimates that $l(d) \cong 3.5 \times d$. Here we renormalized the BCS coherence length, $\xi_0(d) = \xi_0^{bulk} \times T_{c0}/T_c(d)$, to account for the slightly lower $T_c$ in thin films. Evidently, $l(d)$ and, consequently, $\xi_{GL}$ are limited by boundary scattering.

Thus far, the anisotropic GL formalism works surprisingly well in this quantum confined regime. However, $H_{c2}(T)$ deviates significantly from linearity as $T$ approaches $T_c$ from below. While a slight rounding of $H_{c2}(T)$ near $T_c$ may arise from structural inhomogeneities in conjunction with the boundary conditions for the pair wave function[22,23], the rounding observed here changes systematically with the fundamental nanoscale dimension $d$. Extrapolation of the linear $H_{c2}(T)$ segments to zero DC field (Fig. 1) produces a set of extrapolated temperatures $T_c^*(d)$ which also scale linearly with $1/d$ (inset). Accordingly, in the GL regime below $T_c^*(d)$, the slope $-dH_{c2}/dT$ should be proportional to $(d - d_c^*)^{-1}$ as is indeed observed experimentally ($d_c^* \cong 2.4$ ML is the thickness where $T_c^*(d)$ extrapolates to 0 K). One might interpret $T_c^*(d)$ as an extrapolated GL or "mean field" transition temperature. This, however, would imply that the region between $T_c^*(d)$ and $T_c(d)$ is dominated by 2D fluctuations with only a fluctuation conductivity. Instead, as demonstrated below, the $H_{c2}(T)$ phase boundary coincides with the onset of irreversible magnetization, indicating the presence of a robust critical current density above $T_c^*(d)$.

Figure 2(a) - (d) shows scanning tunneling microscopy images of 7 and 9 ML thick Pb films on Si(111), with corresponding DC magnetization loops in (e) – (h). Because we intentionally underdosed or overdosed the amount of Pb, one observes either nanoscale voids, shown in (a) and (c), or nanoscale mesas, shown in (b) and (d). The voids and mesas are *exactly* two atom layers deep or two atoms layers tall, respectively[11]. They are stabilized by strong quantum size effects and constitute quantum growth "defects". The contrast in magnetic properties of the underdosed and overdosed films is striking. In particular, the highly irreversible magnetization of the underdosed films with voids [Fig. 2(e) and 2(g)] implies a near-perfect Bean-like critical state (*i.e.*, a very hard superconductor) with exceptionally strong vortex pinning[20,24]. Using the critical state relation $J_c = 30m/Vr$ where $m$ is the DC magnetic moment, $V$ is the film's volume, and $r$ the macroscopic radius of the sample ($\approx 1.5\,\mathrm{mm}$), we obtain $J_c$ = 2.0 MA/cm$^2$ for the underdosed 9 ML film at 2 K in 5 Oe DC field. In contrast, small DC fields quickly depress the magnetization of the overdosed films [Fig. 2(f) and 2(h)], indicating that vortex pinning with mesas is weak.

A comprehensive set of $J_c(H,T)$ data was obtained by measuring $\chi''$ in various AC+DC fields. Within the critical state model, the peak position of the imaginary AC susceptibility $\chi'' = (m''/h_{AC})$ corresponds to the condition $J_c = 1.03 h_{AC} d^{-1}$ where $h_{AC}$ is the AC modulation amplitude[25,26]. For the Pb films, the onset of $\chi''$ always coincides with the onset of diamagnetic screening (or $\chi'$), indicating that the thermodynamic $H_{c2}(T)$ phase boundary closely coincides with the onset of irreversible magnetization and establishment of a Bean-like critical state. Notice that the $J_c$ obtained in this way does not depend on the sample radius $r$. Hence, the close agreement between $J_c$ values from the AC susceptibility (2.8 MA/cm$^2$) and DC magnetization (2.0 MA/cm$^2$) implies that the critical current paths closely follow the circumference of the sample. *The ~2 MA/cm$^2$ critical currents are truly macroscopic despite the extremely thin geometry and despite the presence of surface steps*. In fact, the $J_c$ of the underdosed films is as high as ~10% of the depairing current density, $J_d \approx H_c / \lambda_{eff}$, which is about 20 MA/cm$^2$ for a 9 ML thin film, an amazing result indeed. Here, we estimated the penetration depth $\lambda_{eff}$ and thermodynamic critical field $H_c$ assuming that, in accordance with the Anderson theorem[20], the product $\lambda\xi = (\lambda_L \xi_0')_{bulk} \approx (\lambda_{eff} \xi_{GL})_{film}$ is independent of scattering and thus independent of the film thickness. Accordingly, $\lambda_{eff}(0) \approx 1500$ Å for a 9 ML film.

Figure 3 shows the temperature-dependent critical currents for 7 and 9 ML films in 5 Gauss DC fields. Again, the contrasting behavior between the voids and mesas is clearly evident. Presumably, the nanoscale voids greatly enhance $J_c$, which must be attributed to strong vortex pinning. Here, we show that the magnitude and temperature dependence of the critical current can be calculated quite precisely from the known geometry of the voids. The scale of vortex line energy *per unit length* is given by[20] $\varepsilon_0 = \Phi_0^2 / (4\pi \lambda_{eff})^2$. Consequently, 2 ML deep voids are effective trapping centers for cores of the Pearl vortices as they significantly reduce their line energy inside the voids. In other words, *nano-voids attract and pin vortices while nano-mesas repel them*. The pinning centers can be viewed as a uniform-depth segment of a columnar defect. A slight modification of Nelson and Vinokur's expression for columnar defects[27] provides the theoretical estimate,

$$J_c = [c\Phi_0 / (4\pi)^2 \lambda_{eff}^2 \xi] \times (\Delta d / d)$$

where $\Delta d$ = 2 ML is the depth of the pinning centers ($c$ is the speed of light). This gives $J_c(T=0)$ of about 4 MA/cm$^2$ and a vortex pinning energy $U_0 = \varepsilon_0 \Delta d$ of ~500 K. The estimated $J_c$ value is remarkably close to the experimental 2.8 MA/cm$^2$. Furthermore, in the GL regime below $T_c^*$, the critical current $J_c \propto (T - T_c^*)^{-3/2}$ (Fig. 3: inset), consistent with the $(1 - T/T_c)^{-1/2}$ variation of $\lambda_{\text{eff}}$ and $\xi$ in GL theory, and the observed $H_{c2}(T)$.

The robustness of the critical state can be quantified in terms of the creep exponent $n$ in the current-voltage relation $E = E_c (J/J_c)^n$ ($E$ is the electric field)[25,26]. Large values of $n$ imply a sharp demarcation between loss-free current flow and dissipative conduction. The Bean critical state corresponds to $n = \infty$ while $n = 1$ corresponds to Ohmic transport. The creep exponent can be deduced by plotting $\chi'$ versus $\chi''$ in a Cole-Cole diagram[25,26]. Fig. 4a shows that the magnetic data of the 9 ML film collapse beautifully, lying very near the theoretical Cole-Cole curve for the Bean critical state. Direct comparison between the data and theoretical Cole-Cole plots for finite creep exponents[25,26] indicate that $n > 100$.

This large $n$-value was verified independently by real time measurements of the current decay rate (Fig. 4b). This experiment was done with the sample stationary[28], and the data were analyzed according to $J(t) = J_0 - J_1 \ln(1 + t/\tau)$ where $\tau$ is an initial transient time[29]. The normalized creep rate $S$ then follows from the relation $S = J_c^{-1} dJ/d \ln(t) \cong J_1/J_0 = T/U_0$ in a Kim-Anderson formulation[20]. This procedure yields a creep exponent $n = 1 + S^{-1}$ of ~100, consistent with conclusions from the Cole-Cole analysis; it gives a pinning energy $U_0 \cong 700$ K, in good agreement with our previous estimate $U_0 = \varepsilon_0 \Delta d \approx 500$ K.

The experiments reveal with considerable precision the thermodynamics and non-equilibrium response of a hard nanoscale superconductor. The surprising robustness follows from the minimal disorder in these high quality films, coupled with efficient quantum trapping of magnetic flux lines. In nanostructures, thickness variations are atomically discrete and comparable in magnitude with the overall thickness or size. Therefore, strong vortex pinning may be realized in many superconducting nanostructures whose size can be controlled with atomic precision, as demonstrated here.

Finally, the applicability of 3D anisotropic GL theory for this extreme 2D geometry may be rationalized on the basis of the time-energy uncertainty principle $\Delta E \geq \hbar/\tau$ where $\tau(d) = l(d)/v_F$; $v_F$ is the Fermi velocity. In this extremely thin limit, $\Delta E$ is at least a few tenths of an eV, which is comparable to the inter-subband spacing of quantum confined Pb[15,17]. Hence, carriers may easily be scattered between 2D subbands so that superconductivity is no longer strictly 2D. Strict two-dimensional Cooper pairing thus requires long mean free paths, meaning specularly reflecting interfaces. It should be possible to push superconductivity closer to this "clean limit" by judicious choices of the substrate and capping layer. Hopefully this work will inspire new efforts toward creating atomically abrupt nanoscale superconductors with nearly perfect interfaces. Such structures not only present an ideal testing ground for theories of low-dimensional superconductivity with quantifiable parameters, but may also unveil novel and unexpectedly robust critical state properties that could be useful for superconductive nano devices.

**Methods:** Pb films were grown[11] in ultrahigh vacuum and protected by an amorphous Ge cap layer. Their superconductive properties were measured inductively as a function of temperature and perpendicular magnetic field, using a SQUID magnetometer. For AC studies, a small 100 Hz AC probing field was superimposed on the DC field. These external fields generate circulating screening currents and an associated magnetic moment *m*. The AC moment contains a diamagnetic in-phase term and a lossy, out-of-phase component, so that $m = m' - im''$.

**Acknowledgment.** This work has been funded by the National Science Foundation under Contract No. DMR-0244570. Oak Ridge National Laboratory is managed by UT_Battelle, LLC, for the US Department of Energy under Contract No. DEAC05-00OR22725.

**Figure legends**

**Figure 1 Equilibrium superconductive properties of ultrathin, atomically flat Pb films.** Main panel shows upper critical fields as a function of temperature. Notice the systematic thickness dependence of the slope $-\mathrm{d}H_{c2}/\mathrm{d}T$, and the rounding and flattening of $H_{c2}$ in the vicinity of $T_C$, particularly for the thinner films. The rounding suggests the existence of a $T_C^*$, as indicated for the 9 ML film. The inset shows superconductive $T_C$ and $T_C^*$ plotted as function of the inverse of the film thickness $d$. The critical temperatures $T_C$ are obtained from the onset of diamagnetic screening, while the values $T_C^*$ are obtained from the extrapolated $H_{c2}(T)$ data, using a 10 mG AC probing amplitude. For $d \to \infty$, $T_C^*$ extrapolates to (7.03 ± 0.13 K), about 1½ standard deviations from the bulk $T_{c0}$ (7.2 K). In the blue shaded region, $\xi$ is limited by boundary scattering and, accordingly, $H_{c2}(T)$ is linear in temperature. The yellow region, where $\xi_{\mathrm{GL}}(T)$ becomes very large, indicates a strongly superconductive regime where the pair function averages over short-range scattering inhomogeneities[22] and becomes increasingly insensitive to them.

**Figure 2 Scanning Tunneling Microscopy images**. Quantum growth defects in 7 ML Pb film consist of either (a) two-atom layer deep voids or (b) two-atom layer tall mesas, in films with small shortage or excess of Pb, respectively. Similar quantum growth defects are evident in 9 ML films, with (c) voids or (d) mesas. The 700x700 nm$^2$ images are obtained with STM. The area density of defects is comparable with the density of vortices in a "matching field" of ~5 kG. The corresponding DC magnetic response of these films, nominally 3×3 mm$^2$, is shown in (e) - (h). Quantum voids produce "hard" hysteresis loops

as shown in (e) and (g), while quantum mesas produce "soft" hysteresis loops shown in (f) and (h). Hard hysteresis indicates strong vortex pinning by the voids.

**Figure 3 Critical current density of 7 ML and 9 ML thick Pb films**. Open symbols refer to films with nanoscale voids; filled symbols refer to films with nanoscale mesas. The inset shows that $J_c \propto (T_c^* - T)^{-3/2}$ for the 9 ML film below $T_C^*$, as expected from the modified Nelson-Vinokur expression for $J_C(T)$, in conjunction with the $(1 - T/T_c)^{-1/2}$ variation of $\lambda_{\text{eff}}$ and $\xi$ in Ginzburg-Landau theory.

**Figure 4 (a) Cole-Cole plot of the real and imaginary diamagnetic susceptibility of an underdosed 9 ML thin film**. The ($\chi'$, $\chi''$) data span a wide range of temperatures (1.8 K to $T_C$) and AC modulation amplitudes (200 to 1100 mG). The DC magnetic field is 5 G. The data collapse perfectly, adjoining the theoretical Cole-Cole curve of the Bean critical state ($n = \infty$) of a thin disk, which is indicated by the solid line. Similar results were obtained in DC fields up to 5 kOe. **(b) Current decay with time.** Solid line is fitted to the theoretical time dependence (given in the text)[29]. The current decay was analyzed within the "flux creep" framework of Anderson-Kim (Ref. 26). The Cole-Cole plot shown in (a) and the current decay rate shown in (b) both indicate a very sharp demarcation between loss-free current flow and dissipative conduction at the upper critical field, and consistently indicate a very large value of the current exponent ($n \geq 100$), meaning very hard superconductivity.

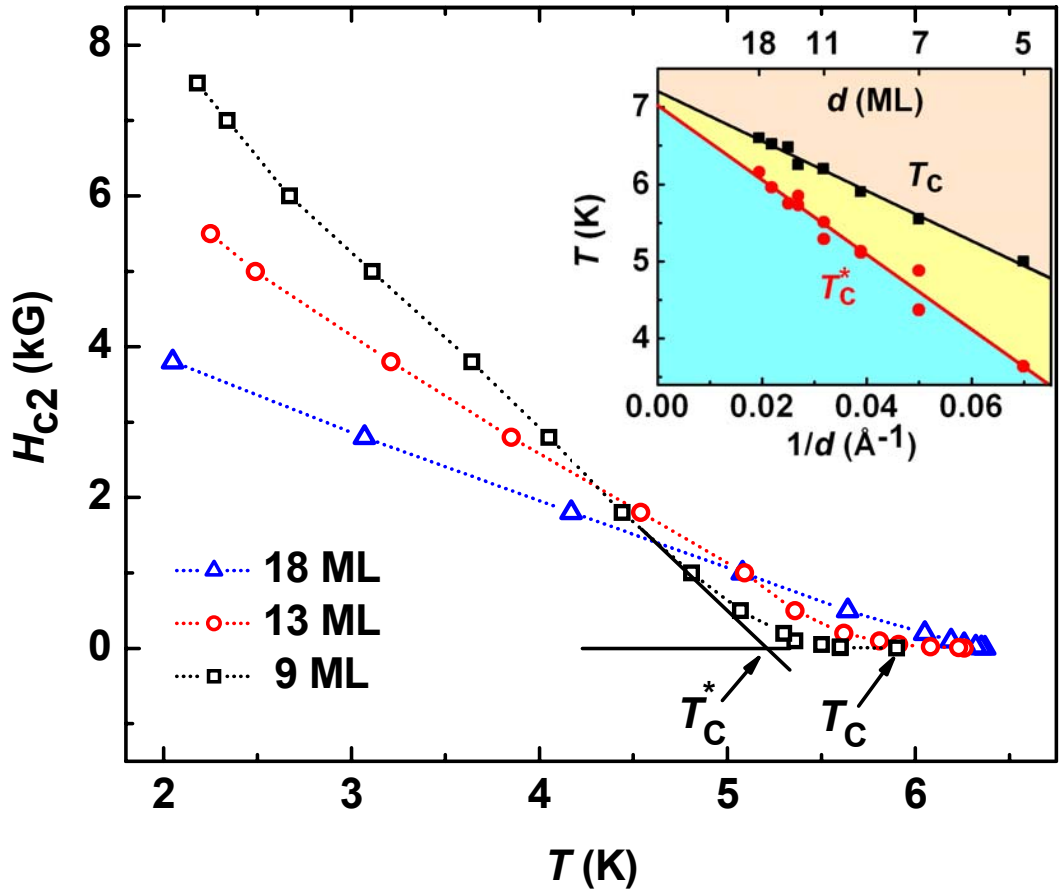

Figure 1

Özer et al

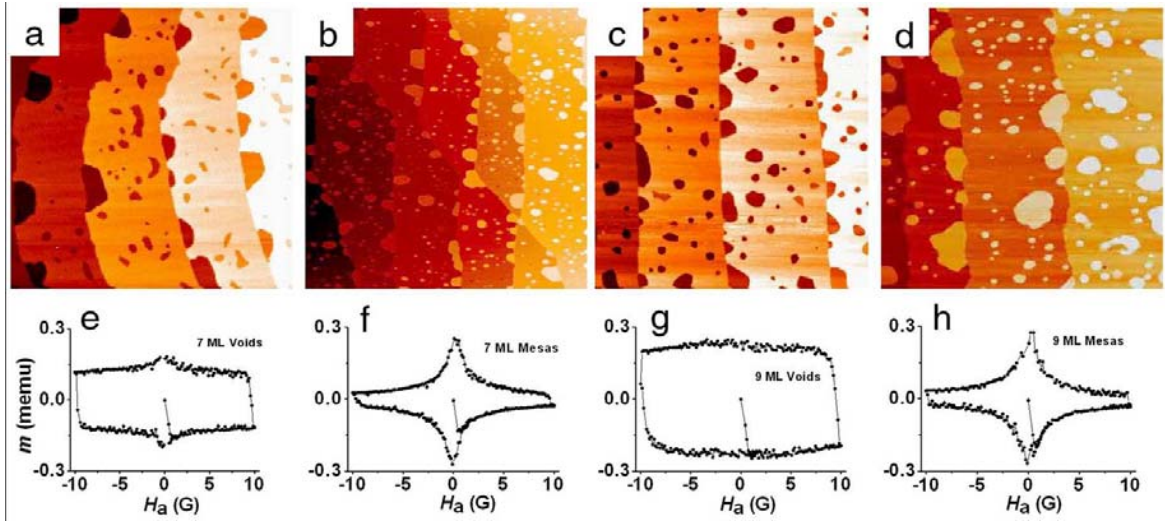

Figure 2

Özer et al

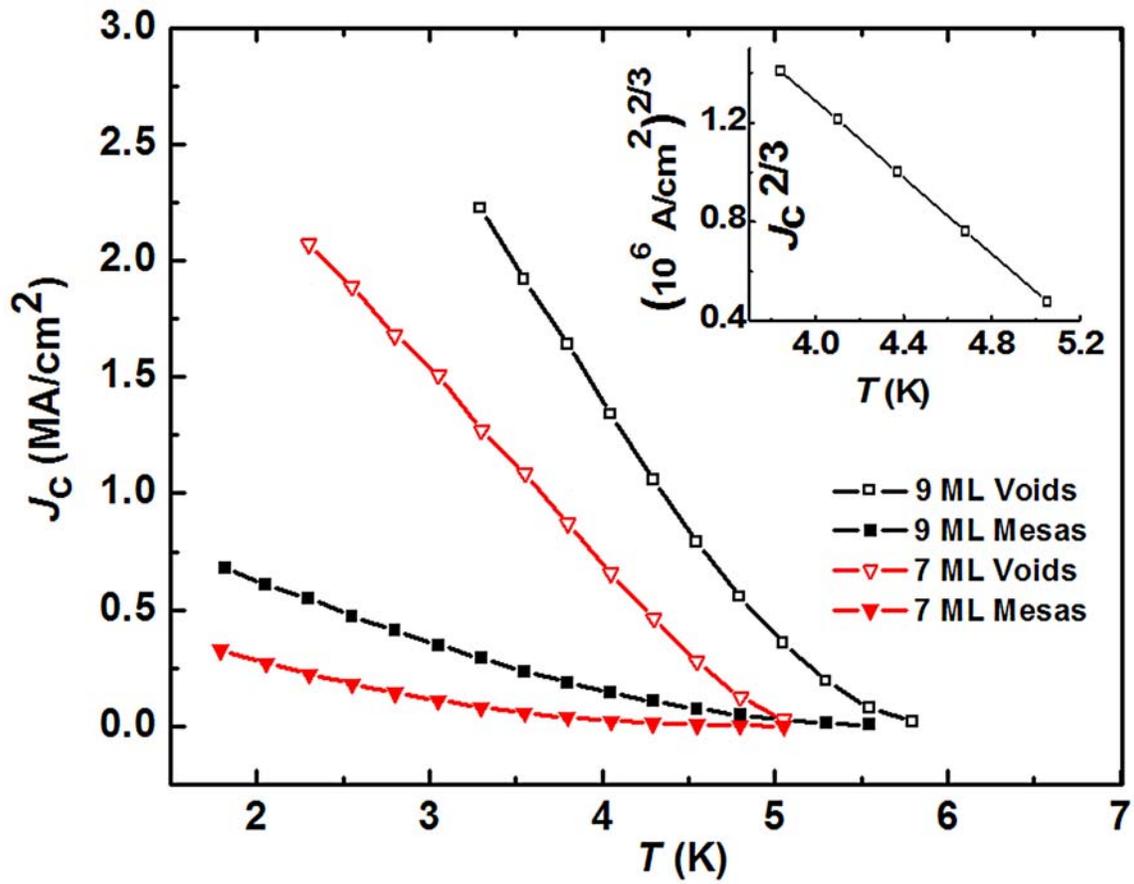

Figure 3

Özer et al

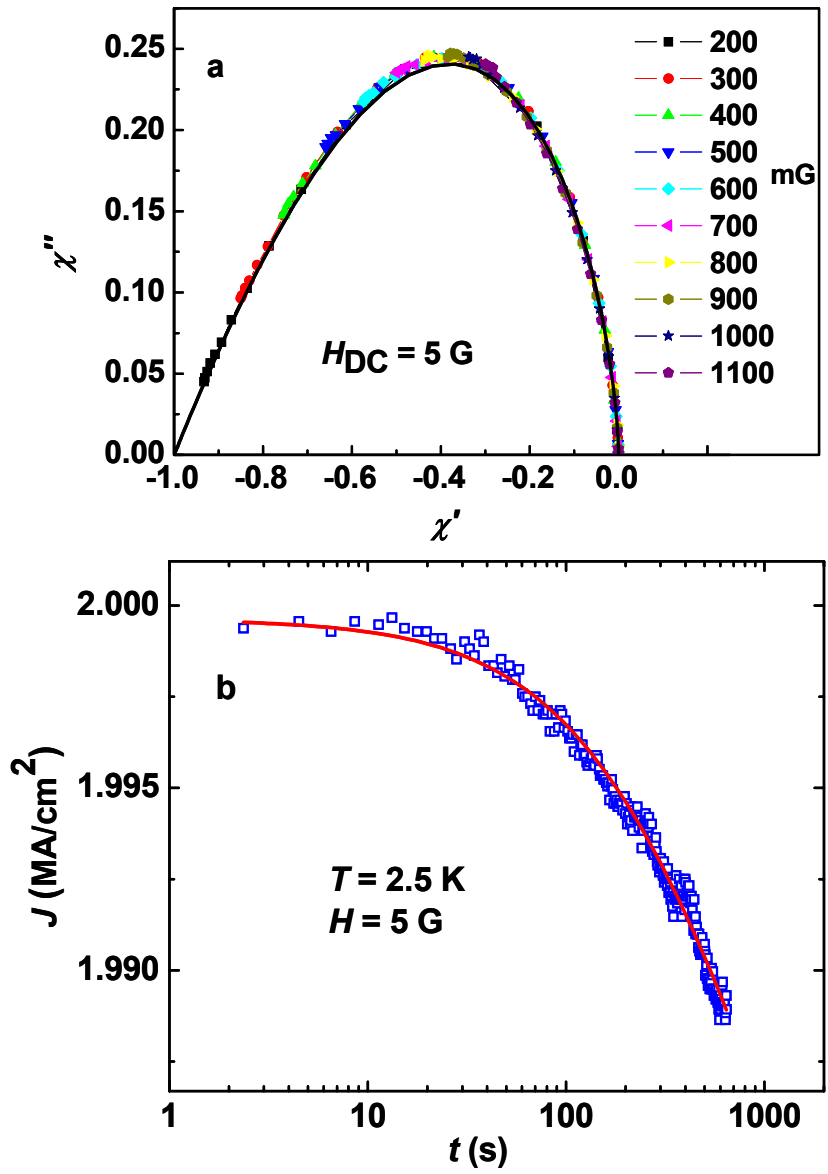

Figure 4

Özer et al